\def\beq{\begin{equation}}
\def\eeq{\end{equation}}
\def\ba{\begin{eqnarray}}
\def\ea{\end{eqnarray}}
\def\non{\nonumber}
\def\half {{1\over 2}}

\def\figloc#1#2{\epsfysize=3in
    \centerline{\epsfbox{fig#1.ps}}
    \centerline{Figure #1}
    {\raggedright\it   #2 }
    \bigskip
    }

\documentstyle[preprint,aps,epsf ]{revtex}
\tightenlines

\begin{document}

\title{
Complex Paths and the Hartle Hawking Wave-function for Slow Roll
Cosmologies
 }

\author{W. G. Unruh${}^*$, Moninder Jheeta${}^\dagger$}
\address{
 CIAR Cosmology and Gravity Program\\
Dept. of Physics\\
University of B. C.\\ 
Vancouver, Canada V6T 1Z1\\
~
 email: ${}^*$ unruh@physics.ubc.ca\\
\quad\quad\quad\quad${}^\dagger$ mj227@damtp.cam.ac.uk}

\maketitle

\begin{abstract}
A large set of complex path solutions for the Hartle Hawking semi-classical 
wave function are found for an inflationary universe in the ``slow roll" regime. 
The implication of these for the semi-classical evolution of the universe is also
studied.
\end{abstract}

Hartle and Hawking\cite{HH} suggested that one could overcome the problem of
supplying initial conditions for the universe by supplying a theory of those
initial conditions within the context of quantum gravity in the path integral
formalism. One of the difficulties with any quantum or classical system is that
the theory only predicts the evolution of a system-- ie the correlations between
the states of the physical attributes of a system at different times.
Unfortunately in studying the universe as a whole, there exists no way of
determining what the early state of the universe other than by observations made
now. There is no way of testing whether or not one's theory of the early
universe is actually correct, only of displaying what initial conditions must
have been present at that early time to produce the universe as it is now. Since
any physical theory always has a set of initial conditions which can produce any
arbitrary final state, one is left with aesthetic judgments as to whether or
not the required initial conditions are pleasing or not.

However, within the context of the path integral formulation of quantum Einstein
gravity, Hartle and Hawking demonstrated that one could generate a natural
theory of a replacement for those initial conditions. Namely, since Einstein's
theory allows changes in the  structure of space-time, one could imagine a
requirement on the universe that it not have any initial conditions. The
universe can simply run out of time into the past, with some unique structure
demanded at the time that the universe came into existence. 

This suggestion is not tenable within the context of ordinary General Relativity
because of the various singularity theorems for Lorentzian space-times. However,
for a Euclidean space-time, Einstein's equations allow singularity free solutions
with cross sections representing various possibilities for the universe at some
instant in time. Such solutions, while of dubious relevance for the classical
theory, can come into their own in the quantum theory, especially the quantum
theory formulated in terms of path integrals.

The Hartle Hawking prescription was thus that one could calculate a complex
valued function, called the wave-function of the universe, within the path
integral formalism. This function is a function of three geometries--
interpreted as the geometry of the universe at some particular time-- together
with configurations of classical fields on that that three geometry. The
definition they suggested was that this wave function was to be calculated by
means of a path integral in which one integrated over all non-singular four 
geometries which
had only this three geometry as a boundary, together with fields on that 4
geometry such that the value of those fields on the specified three geometry
were the given fields. I.e.,
\beq
\Psi( {\cal G}^{(3)}), \Phi(x\epsilon {\cal G}^{(3)}) =
\int  e^{iS({\cal G}^{(4)}, \phi(x\epsilon {\cal G}^{(4)}))} \delta
\phi\delta{\cal G}
\eeq
where the four-geometry ${\cal G}^{(4)}$, and the fields $\phi$ defined on that 
four-geometry, are non-singular.

While this formulation of the wave-function is certainly appealing intuitively,
it suffers from numerous technical problems. Since path integrals even in
standard situations are typically taken over highly pathological paths (eg for a
free particle over paths which are continuous but not differentiable), the
requirement that ${\cal G}^{(4)}$ and $\phi$ be non-singular is difficult to specify
exactly. Furthermore, this path integral cannot be evaluated even in very simple cases.
Finally, once one  has calculated this wave-function, it is unclear how it
should be used to actually derive predictions from the theory.

 However the
purpose of this paper is not to address these difficulties. Rather we will take
a standard naive approach to the problem. We will assume that the wave-function
is well approximated by the semi-classical approximation, in which one assumes
that the extrema of the action, $S$ will dominate the path integral.
Furthermore, we will neglect the ``determinant" terms, which arise from taking
into account the   integral over paths which are ``near enough" to the classical
path that one can make the quadratic approximation to the action of these paths
near the semi-classical path. Finally, we will interpret this wave function as
giving us a semi-classical Hamilton Jacobi function, which we can then use to
predict the semi-classical evolution of the universe.

One's first reaction is that surely this semi-classical evolution will be simply
some classical solution to Einstein's equation. The difficulty arises in that
there exist no classical Lorentzian solutions to Einstein's equations which are
non-singular and have only the given three-geometry plus matter fields as a
boundary.  The proposal made by HH was therefore that these semi-classical
solutions be taken to be a combination of Lorentzian and Euclidean four
geometries, matched together at some time. The fiducial example is if we take
the simplest situation, namely one in which the only source for the
gravitational field is a cosmological constant, and that the geometry be
restricted to a maximally symmetric geometry, with three geometries being metric
three-spheres. In this case one can join a De-Sitter Lorentzian universe to a four
dimensional metric sphere (a Euclidean solution to the Einstein equations  with
cosmological constant).
In order that Einstein equations also be valid across the junction between the
solutions it is necessary that the extrinsic curvatures of the three geometries
at the join be continuous. However, given some definition of the time, one can
characterize the Euclidean solution as having imaginary time, while the
Lorentzian has real time. The extrinsic curvature in the Euclidean space is
therefore imaginary while it is real in the Lorentzian, and thus matching can be
smooth only if the extrinsic curvature is zero\cite{GH}. Thus the join between the two
solutions must take place at the minimum 3-volume for the De-Sitter space, and
the maximum 3-volume for the Euclidean-- ie one must match the center of the
De-Sitter throat to the equator of the Euclidean sphere.

This four geometry has the feature that at any ``time" the three geometry
obtained by taking an equal time cross section is a real three geometry, and the
vague idea has developed that this is typically what one wants to have happen.
However, one begins to quickly realize that, in general (ie in more
complicated situations), this is impossible. Essentially, the requirement that one  have
a real three-geometry at each and every time would require a matching between
the Euclidean and Lorentzian regimes such that the momenta corresponding to
each of the fields, and to the geometry would all   be zero at the
junction. If not, the momenta would have to be complex in one or the other of
the regimes, and would thus quickly (because of the equations of motion) create
complex field strengths and complex metrics. However, as we shall see in   
very simple cases, there exist no solutions to the classical Euclidean  Einstein
equations which are both regular and have one three boundary on which all of
the momenta of all of the fields and of the geometry are zero.

One is therefore driven to investigating complex three-geometries, and complex
four geometries as classical solutions to Einstein's equations. (We also advise
the reader to read the papers by Halliwell and Louko \cite{HL} and Hartle and
Halliwell\cite{HartH} which contain a deeper discussion of the motivation behind
finding complex solutions. Note also the paper by G. Lyons \cite{Lyons} who qualitatively
examined the case where the potential for $\phi$ is quadratic and arrives 
at some of the same
conclusions we do). In this paper we
will investigate the complex solutions for a simple problem, namely one in which
most of the degrees of freedom of the geometry and the matter fields are frozen, so that
the geometry has a    high  degree of symmetry (in particular
having the symmetry of the metric three-sphere) but in which there exists a
matter field given by a scalar field with a non-trivial potential, but with the
same spherical symmetry of the metric. The potential for the scalar field will
be assumed to be almost flat (ie with small slope) and we will examine the
complex solutions as a power series in that slope of the potential. With zero
slope this problem of course simply reduces to that originally studied by Hartle
and Hawking, since a flat scalar field potential is equivalent to  a
cosmological constant.

We will find that there are in fact an infinite number of complex semi-classical
solutions. Many of them will be analytically related to each other. They will also give
an action for the final state of the system which is the same for large classes of the 
paths. There are however an infinite number of analytically inequivalent paths,
many of which lead to different wave-functions. 

Furthermore, if we examine the classical solutions implied by these solutions,
we find that the classical space-time need not be the same as (in fact because of
the complex nature of the extremal solutions, cannot be the same as) any of the
extremal paths. We suggest that one interpretation of the Hartle Hawking wave
function is that the classical universe has no initial singularity because it
bounces. Ie, we suggest that the Hartle Hawking wave function is a way of
imposing the requirement on the quantum system that the universe   bounce (
perhaps due to quantum corrections) before it hits a singularity.

\section{The model}

The universe is assumed to have the geometry
\beq
ds^2= N(t)^2 dt^2 - a(t)^2 \left(dr^2+sin^2(r)(d\theta^2 +sin^2(\theta) 
d\varphi^2)\right)
\eeq
with a homogeneous scalar field $\phi(t)$. Since we will be interested in
complex extrema to the action, we will assume that all of $N$ the lapse
function, $a$ the scale factor, and $\phi$ the scalar field can be complex.
However, we will assume that the real universe has real scale factor $a$
 and real
scalar field $\phi$. Thus the wave function of the universe is a function only
of real $a=a_f$ and real $\phi=\phi_f$. Ie, we will calculate $\Psi(a_f,\phi_f)$
via the semi-classical approximation.

The action will be assumed to be the standard Einstein Hamiltonian action,
\beq
S=\int\left(\pi^{ij} \dot\gamma_{ij} +\pi_\phi \dot\phi - N H_0 -N_i H^i
\right) d^3x dt
\eeq
where
\beq
H_0=  \left({1\over \sqrt{\gamma}}\pi^{ij}\pi^{kl}\left(\gamma_{ik}\gamma_{jl} - \half
\gamma_{ij}\gamma_{kl}\right) -\sqrt{\gamma} R+ (\half{\pi_\phi}^2
+\sqrt{\gamma}({1\over 2\sqrt{\gamma}}\phi_{i}\phi_{j}\gamma^{ij} + V(\phi)\right)  
\eeq
\beq
H_i= \pi^{ij}_{,j} +\Gamma^i_{jk}\pi^{jk} +\pi_\phi \phi_j \gamma^{ij}
\eeq
Because of the symmetry we have assumed for our fields, $H_i$ is automatically
equal to zero, and we can write the action in terms or $a$, $\phi$ and their
conjugate momenta $\pi_a$ and $\pi_\phi$ so that 
\ba
S= \int \pi_a\dot a +\pi_\phi \dot\phi - N H dt\\
H= -{1\over 24 a}\pi_a^2 - 6a
 + ( {1\over 2a^3} {\pi_\phi}^2 + a^3 V(\phi))
\ea

The Hartle Hawking procedure is often characterized as  allowing time to become
imaginary. However, we will assume that $t$ is real throughout. The transition to
Euclidean space-time will be done by allowing $N$ the lapse function to become purely
imaginary. However, in addition to allowing $a$ and $\phi$ to become complex (as
they must) we will also allow $N$ to be not only purely real or purely imaginary
but also to be an arbitrary complex function of the real time $t$. Ie, we will be
looking at complex four-metrics and complex
fields on a real four dimensional manifold characterized by coordinates  $t~ r~
\theta~ \varphi$.

Despite the simplicity of this action, the extrema are still 
difficult to find analytically. We will therefore find them by assuming that
$V(\phi)$ is almost constant, and can be written as
\beq
V(\phi)= V_0 +\epsilon  \phi 
\eeq
We will then assume that the solution is analytic in $\epsilon$ and
 solve our problem as a power series in $\epsilon$. Ie, we assume that $V(\phi)$
 is essentially that for inflation, with the solutions we examine being for
 conditions such that a significant era of inflation takes place.

It will be useful for future work to note that this action has a symmetry. In
particular if we allow $V_0$ to go to $V_0+\epsilon Q$ and $\phi$ to go to
$\phi-Q$ the action is left invariant. 

The extrema are defined by the equations
\ba
{1\over N} {da\over dt} &&= -{1\over 12a}\pi_a\\
{1\over N} {d\phi\over dt}&&={1\over a^3}\pi_\phi\\
{1\over N} {d\pi_a\over dt}&&=-\left[ {1\over 24a^2}\pi_a^2-6 -{3\over 2a^4}
\pi_\phi^2+3a^2V(\phi)\right]\\
{1\over N} {d\pi_\phi\over dt}&&= -a^3 {dV\over d\phi}
\ea
We also have the fifth equation, an integral of the motion of the above
equations and the constraint equation corresponding to the variation of
the action with respect to $N$, of
\beq
H=\left[ -{1\over 24 a}\pi_a^2 - 6a
 + ( {1\over 2a^3} {\pi_\phi}^2 + a^3 V(\phi))\right]=0
 \eeq
 Using the equations of motion, this can be written as
 \beq
-6aa'^2 -6a +\half a^3\phi'^2 +a^3V(\phi)=0
\eeq
where $'$ denotes ${1\over N}{d\over dt}$.
We note that N occurs in these equations only in the combination $Ndt$. We can
thus define a new variable $\tau$ by
\beq
\tau=\int N(t)dt
\eeq
where we will take the zero of $\tau $ to occur at the ``time" when $a$ is zero.
Since we are allowing $N$ to be complex, $\tau $ will also be a complex function
of $t$. $\tau(t)$ will be a parameterized path through the complex $\tau$ plane.
Note that these different paths are not simply different coordinates. A
coordinate transformation on $t$ cannot create a change in for example the ratio
of the real to imaginary part of $\tau$. The paths through $\tau$ will be
coordinate invariant paths. $'$ will then denote the derivative with respect to
$\tau$.

We want the four geometry to be non-singular at $\tau=0$. Since $\tau=0$ is
defined to be the initial point along the path where $a=0$, the regularity
condition on $a$ will be one on  the derivative of $a$ and on the momenta $\pi_a$
and $\pi_\phi$. Examining the metric, we see that it is regular at $a=0$ if at
that point $a$ goes as $\tau$ and $\tau$ is purely imaginary so as to create a
Euclidean metric. (For example, the curvature, $G_0^0 = 3(a'^2+1)/a^2$, is
non-singular  as $a$ goes
to zero only if $a'^2+1$ goes to zero at least as fast  as $a^2$ does) Thus, the boundary condition at $\tau=0$ is that $a'={da\over
d\tau}=\pm i$ at $\tau=0$. Furthermore, the boundary condition of $\pi_\phi$ is that
$\pi_\phi$ must go to zero, or the equation for $\pi_a$ cannot drive it to zero
sufficiently quickly at $\tau=0$. We thus have the conditions
\ba
a(0)&&=0\\
a'(0)&&=\beta i \Rightarrow \pi_a= -12\tau {\rm ~as~ }\tau\rightarrow 0\\
\pi_\phi(0)&&=0
\ea
where $\beta=\pm 1$.
We also have the final conditions
\ba
a(\tau_f)=a_f=\pm |a_f|\\
\phi(\tau_f)=\phi_f
\ea
where however $\tau_f$ is a free variable which can be chosen so as to allow
these equations to be obeyed. As expected we thus have five boundary conditions
on the two  variables obeying second order equations, $a,\phi$ and the third
variable $\tau_f$. Both signs for the final radius $a_f$ are valid since it is
only $a^2$ which enters the metric.

The action in terms of the $\tau$ variable becomes
\beq
S=\int_0^{\tau_f}\pi_aa'+\pi_\phi \phi' -H  d\tau.
\eeq
We will solve the classical equations of motion in the  $\tau$ variable and as a
power series in the slope of the linear potential, $\epsilon={dV(\phi)\over
d\phi}$. At each order in $\epsilon$, the point $\tau_f$ at which $ a(\tau_f)=a_f$
will also change. We will define $\tau_0$ as the value of $\tau_f$ for
$\epsilon=0$. If $q_i$ is one of our dynamic variables, $a$ or $\phi$ and if $q_i^{(0)}$ is the
zeroth order solution, and 
\beq
q_i=q_i^{(0)} +\sum {\epsilon^n\over n!} \delta^n q_i
\eeq
Then
\beq
q_i(\tau_f(\epsilon))=q_{if}
\eeq
gives
\ba
\delta q_i(\tau_0) &&= -\delta \tau_f {q'_i}^{(0)}(\tau_0) \\
\delta^2q_i(\tau_0)&&= -\delta^2 \tau_f {q'_i}^{(0)}(\tau_0)
 -(\delta\tau_f)^2{q''_i}^{(0)}(\tau_0)-2\delta\tau_f\delta q'(\tau_0)
\ea
We shall use these equations for $q_i=a$ to calculate the expansion of $\tau_f$.

Let us now examine the expansion of the action, 
\beq
S=\int_0^{\tau_f}(\pi_i q_i' -H)d\tau
\eeq
To zeroth order we have
\beq
S^{(0)}=\int \pi_a^{(0)} {a^{(0)}}' d\tau
\eeq
We also  have, for arbitrary $\epsilon$,
\ba
{dS\over d\epsilon} = &&{d\tau_f\over d\epsilon}\left[
\pi_i(\tau_f(\epsilon),\epsilon)q_i(\tau_f(\epsilon),\epsilon)'
-H(\pi_i,q_i,\epsilon)\right] \non\\
&&\quad \quad +\int_0^{\tau_f} {\partial \pi_i\over \partial\epsilon} q'_i
+\pi_i{\partial q'_i\over \partial \epsilon} -\left({\partial H\over
\partial\pi_i}{\partial\pi_i\over\partial\epsilon} +{\partial H\over
\partial q_i}{\partial q_i\over\partial\epsilon} +{\partial H\over
\partial\epsilon}\right)d\tau\\
=&&{d\tau_f\over d\epsilon}\left[
\pi_i(\tau_f(\epsilon),\epsilon)q_i(\tau_f(\epsilon)'\right]
 -\left. \left(\pi_i{\partial q_i\over
\partial\epsilon}\right)\right\vert_0^{\tau_f} - \int_0^{\tau_f} {\partial H\over
\partial\epsilon}d\tau
\ea
where we have used the Hamiltonian equations of motion and the constraint
equation $H=0$ to simplify the equations. But, 
\beq
{d\tau_f\over d\epsilon}q'(\tau_f(\epsilon),\epsilon) +{\partial
q_i(\tau_f(\epsilon),\epsilon)\over\partial\epsilon} = {dq_{if}\over d\epsilon}=0
\eeq
so we finally have
\beq
{dS(a_f,\phi_f,\epsilon)\over d\epsilon}
=-\int_0^{\tau_f} {\partial H(\pi_i,q_i,\epsilon)\over
\partial\epsilon} \label{epsilon}
\eeq

In our case, we have that 
\beq
{\partial H\over \partial\epsilon}= a^3\phi
\eeq
But, we also have from the equation of motion for $\phi$ that
\beq
a^3 = -{1\over \epsilon}\left(a^3\phi'\right)'
\eeq
Thus
\ba
{dS\over d\epsilon} &&= {1\over\epsilon}\int_0^{\tau_f}
(a^3\phi')'\phi d\tau\\
&&= {1\over \epsilon}a_f^3 \phi(\tau_f)' \phi_f - {1\over\epsilon}\int_0^{\tau_f} a^3(\phi')^2
d\tau  \\
&&={1\over\epsilon}\phi_f \int_0^{\tau_f} (a^3\phi')' d\tau - {1\over\epsilon}\int_0^{\tau_f} a^3(\phi')^2
d\tau\\ 
&&= -\phi_f\int_0^{\tau_f} a^3 d\tau - {1\over\epsilon}\int_0^{\tau_f} a^3(\phi')^2
d\tau\\ 
&&=-\phi_f\int_0^{\tau_f} {\partial H\over \partial V_0}d\tau - {1\over\epsilon}\int_0^{\tau_f} a^3(\phi')^2
d\tau 
\ea
However, by exactly the same line or argument which led to eqn \ref{epsilon},
we have
\beq
{\partial S(a_f,\phi_f,V_0,\epsilon)\over \partial V_0} = -\int_0^{\tau_f}{\partial H\over
\partial V_0} d\tau
\eeq
Thus we finally have
\beq
{dS\over d\epsilon}= \phi_f{\partial S\over \partial V_0} 
- {1\over\epsilon}\int_0^{\tau_f} a^3(\phi')^2
d\tau 
\eeq
We can solve this in part by requiring that $V_0$ occur in $S$ only in the combination
$V_0+\epsilon\phi_f$. 

For our particular problem, the second term on the right is of order $\epsilon$.
Thus, the first order term $\delta S$ is determined by the zero order term, and
the first interesting terms is the second order term $\delta^2S$. Since
$\phi'=\epsilon \delta \phi ' + O(\epsilon^2)$, the second order term in $S$ is
determined by the zeroth order term in $a$ and the first order term in $\delta
\phi$.
 
We will  look at  the first order solutions for both
$a$ and $\phi$, namely $\delta a$ and $\delta\phi$.

\section{ Solutions and Actions}
\subsection{Zeroth Order solutions}

With our potential, $V(\phi,\epsilon)=V_0+\epsilon\phi$ we have the equations of motion 
\ba
-6aa'^2 -6a+\half a^3\phi'^2+a^3 (V_0+\epsilon\phi)&&=0\\
(a^3\phi')'&&=-a^3\epsilon
\ea
The zeroth order solutions give that $a^3\phi'$ a constant. If this constant 
were non-zero,
then $\phi$ would diverge at $\tau=0$ at least as $1/\tau^2$ and the solution
would not be regular. Thus, to zeroth order, $\phi$ is a constant, and from
the final boundary condition $\phi(\tau)=\phi_f$.

The zeroth order solution for $a$ of 
\beq
{a^{(0)}}'^2 = a^{(0)2}(V_0/6)-1
\eeq
is (with the appropriate boundary condition)  
\beq
a^{(0)}(\tau)= \beta\sqrt{6\over V_0} \sin(i\sqrt{V_0\over 6}\tau)
\eeq
We will define $\nu=\sqrt{V_0\over 6}$, since it will occur time and again, so
the solution   becomes
\beq
a^{(0)}(\tau) = \beta{1\over\nu} \sin(i\nu \tau)
\eeq
This solution obeys the boundary condition on $a$ at $\tau=0$. We also have the
boundary condition
$a^{(0)}(\tau_0)= a_f$. We readily see that this equation has a number of
solutions, since $\sin(i\nu\tau)$ is periodic in imaginary $\tau$ with period
$2\pi i/\nu$. Since $a_f$ is real, we have, defining
\ba
\tau_R&&=Real(\tau_0) \\
\tau_I&&=Imag(\tau_0) \non
\ea
that $\tau_R$ and $\tau_I$ obey
\ba
\beta \cosh(\nu\tau_R)\sin(\nu\tau_I) =  \nu a_f\\
\cos(\nu\tau_I)=0
\ea
For $a_f>1/\nu$, this  has solutions for $\tau_I$
\beq
\tau_I= ( 2n+ {\rm sign}(a_f) \beta\half) {\pi\over\nu}
\eeq
for all $n$. 
In addition, the equation for $\tau_R$
\beq 
\cosh(\nu\tau_R)= \nu |a_f|
\eeq
has  both positive or negative solutions for $\tau_R$. In what follows an
important variable will be $\cos(i\nu\tau_0)$ whose sign depends on both  the
sign of $a_f$ and on the sign of $\tau_R$. I will define the sign, $\gamma$, (ie,
$\gamma=\pm 1$) by
\beq
\cos(i\nu\tau_0)=\gamma \left(+i \vert\sqrt{\nu^2 a_f^2 -1}\vert\right)
\eeq
 for $\nu a_f>1$. In this case 
 \beq
 \gamma= {\rm sign}(\tau_R){\rm sign}(a_f)
 \eeq
 For $\nu a_f<1$, 
 $\tau_R$ must be zero, and $\tau_I$, the imaginary part
of $\tau$ must obey
\beq
\sin(\nu\tau_I)=-\beta\nu a_f
\eeq
and $\gamma$ is given by
 \beq
 cos(i\nu\tau_0)= \cos(\nu\tau_I)=\gamma \left(+  \vert \sqrt{1-\nu^2 a_f^2  }\vert\right).
 \eeq
 In most of what follows we will assume that we are dealing
 with $\nu a_f>1$.

Note that for $\tau_R=0$ and $\tau_I=m\pi/\nu$, $a$ goes to zero.   $m=0$ corresponds
to the initial value $\tau=0$. For even $m$, these zeros will be repetitions of
the initial state. However, for 
    odd $m$ these
zeros will turn out to be singularities in the solutions of the 
equations at higher orders of $\epsilon$.

\figloc{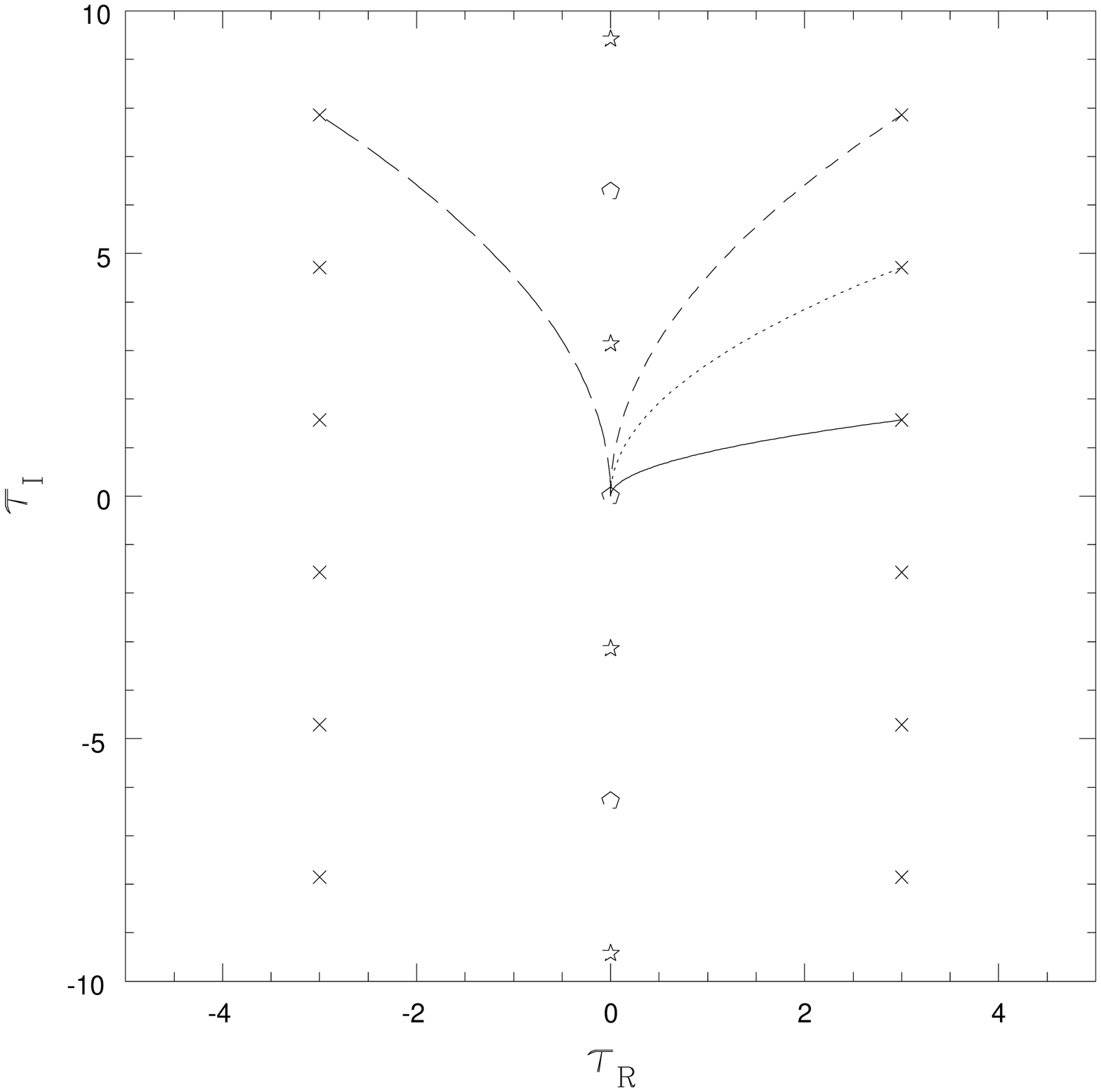}{The location of the zeros of $a(\tau)$ and various possible endpoint values for a given $a_f$ of the
paths in complex $\tau$ space.}

Figure 1 shows these poles, the points where $a $ is zero and values of $\tau_0$
which satisfy the $|a^{(0)}(\tau_0)|= |a_f|$.  in the complex $\tau$ plane. 
This figure also also
indicates a few possible paths that the complex function $\tau(t)$   could
follow.

\subsection{Zeroth Order action}

The zeroth order action is just
\ba
S(a_f,\phi_f)&&=\int_0^{\tau_0} \pi^{(0)}_a {a^{(0)}}' d\tau 
= -\int_0^{\tau_0} 12a^{(0)}({a^{(0)}}')^2 d\tau \\
&&= {24\beta  \over V_0}\gamma\left(  {V_0a_f^2\over 6} -1\right)^{3\over
2}-i{24\beta\over V_0}
\ea

There are many different endpoints $\tau_f$ which give the same $a_f$, and many
different contours to each of these end points. We note that these contours and
end points are ``physically" different in that they correspond to different
paths through the complex $a$ space. Any one of these solutions can be graphed
as a path through the complex $a$ plane. That path is an invariant, and no
coordinate transformation can alter that path. They thus represent physically
distinct paths. However, except for the   parameters $\beta$ ,
$\gamma$ and $sign(a_f)$ differentiating the different solutions, the actual value of the action
as a function of $\phi_f$ and $a_f$ is independent of which of the paths or of
the endpoints are chosen.

In the semi-classical wave function, the real part of the action determines the
phase of the wave-function, while the imaginary part  determines the amplitude.
The two values of $\gamma$ correspond to taking the complex conjugate of the
$iS$ while the two values of $\beta$ correspond to two different values for the
amplitude as well. These two possible values  for the amplitude,
 given by the two possible values of $\beta$, 
 have been the subject of immense controversy. We will not take part in
this controversy, but will keep $\beta$ in the remaining equations. 
We do however note that the issue is not one of determining whether or not the ``current" emitted by the $a=0$ ``singularity" is purely outgoing or not. For
both signs of $\beta$ and $\gamma$ the system obeys the HH condition on the
regularity of the wave-function at $a=0$. Similarly, by taking an appropriate
combination of the wave-functions with $\gamma=\pm 1$ one can have a wave
function with purely outgoing flux, purely in-going or a combination of the two.

The usual path in $\tau$ space, originally chosen in HH, is the path
 through 
 $\tau$ space beginning as a  
purely imaginary path from 0 to $-i\pi/2\nu$ and then going parallel to the real
axis to the endpoint $\left(cosh^{-1}(\nu a_f)/\nu,-i\pi/2\nu\right)$.  
 $a^{(0)}(\tau)$ is real along the whole path. Along any other path, $a$ has an
imaginary part. We note that this path also corresponds to only one of the
possible end values $\tau_0$.  Other end points  correspond to   
trajectories in the complex $a$ plane which wrapping around
$a=0$ more. In figure 2 we plot the trajectories in the complex $a$ plane which
correspond to   the $\tau$ trajectories of figure 1.

\figloc{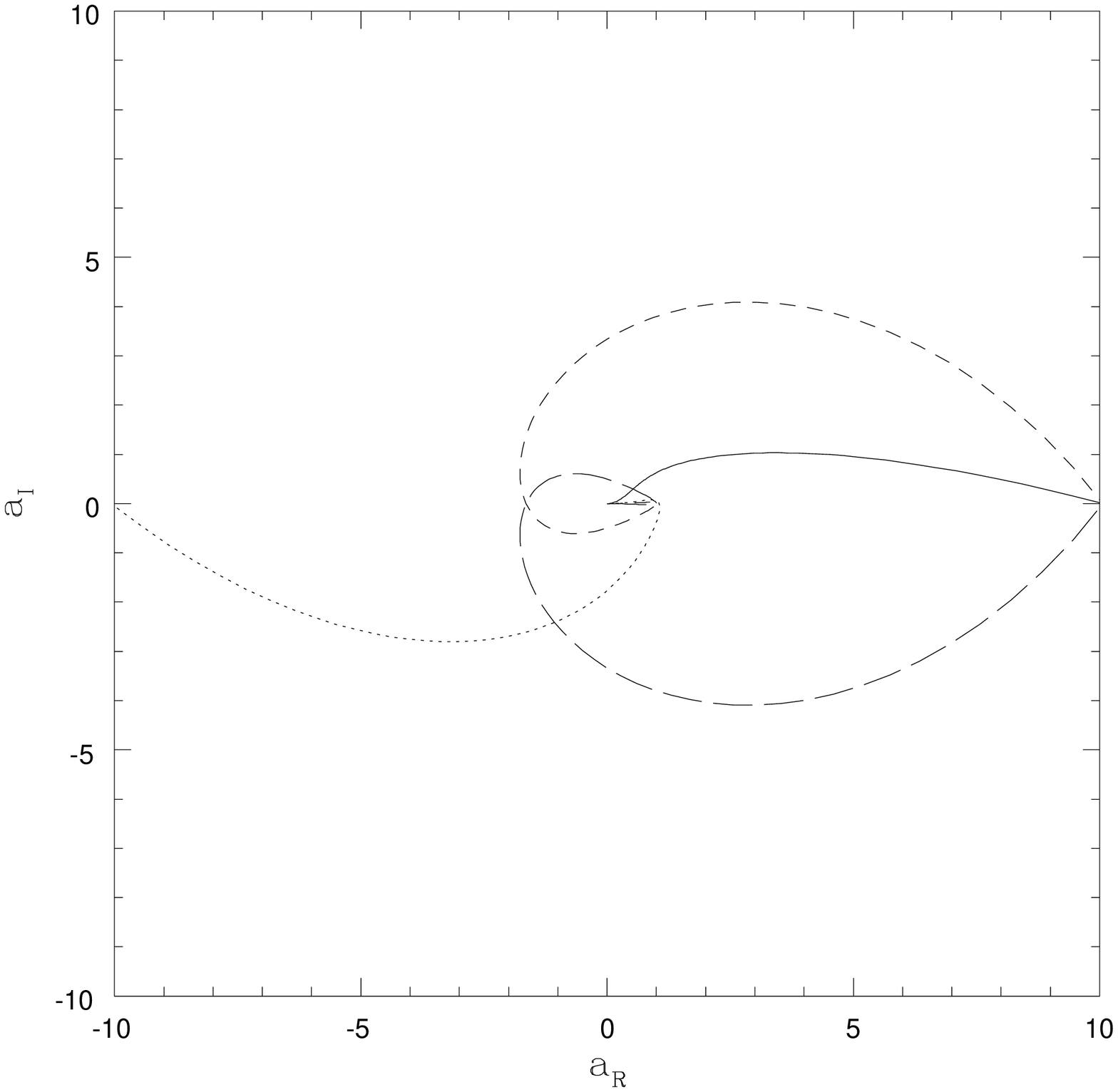}{ The paths in complex scale factor ($a$) space of solutions for the
various complex time ($\tau$) paths of figure 1.}

The function, $S_0(a_f,\phi_f)$ is a Hamilton-Jacobi function for the
$\epsilon=0$ system,
 in that it obeys the equation
 \beq
 -{1\over 24a}\left(\partial S_0\over \partial a\right)^2 -6a +\left({1\over
 2a^2}{\partial S_0\over \partial\phi}\right)^2 +a^3V_0=0.
 \eeq
 which is the Hamilton-Jacobi form of the constraint equation for $\epsilon=0$.

 The first order action is just given by
\ba
 \delta S(a_f,\phi_f,V_0)&&=  {\partial S_0(a_f,\phi_f,V_0)\over \partial V_0}
\ea
This again solves the Hamilton-Jacobi function to first order 
\beq
{1\over 12a}{\partial S_0(a\phi)\over \partial a}
     {\partial\delta S(a,\phi)\over \partial a} +a^3\phi=0
\eeq

\subsection{ First Order Solutions}

The first order equations for $a$ and $\phi$ are required to find the second
order action. Actually, only the first order solution in $\phi$ is needed, but
we will also solve the first order equations for $a$ for future reference. The
equations are
\ba
-12{a^{(0)}}' \delta a' +2V_0 a^{(0)} \delta a +a^{(0)2} \phi_f=0\\
(a^{(0)3}\delta\phi')'=-a^{(0)3}
\ea
with the boundary conditions that all of the variations are zero at $\tau=0$.
Because $\phi^{(0)} $ is a constant, we also have that $\delta\phi(\tau_f)=0$.

The solution for $a$ is 
\ba
\delta a=\beta{\phi_f \over 2V_0} \left(i\tau \cos(i\nu\tau)-{1\over
\nu}\sin(i\nu\tau)\right)
\ea
while for $\delta\phi$ we get first that 
\ba
\delta\phi'&&= -{1\over a^{(0)3}}\int _0^\tau a^{(0)3} d\tau\\
  &&={ {i\over \nu}\left( {1\over
  3}\cos^3(i\nu\tau)-\cos(i\nu\tau)+{2\over 3}\right)  \over \sin^3(i\nu\tau)
  }
\ea
This expression has poles of order 3 at the points $\tau={i\over \nu} (2r+1)\pi$

Note that at the end point, $\tau=\tau_0$, the expression for 
$\delta a$   not only depends on the value of $a_f$, but also on which
particular end point $\tau_0$ is chosen. 
 This implies also that $\delta\tau= {d\tau_f\over
d\epsilon}=- \delta a(\tau_0)/a^{(0)}(\tau_0)'$ will depend on which end 
point is chosen. Although neither of these enter into the expression for the
second order action, we do expect them to have an effect on the third order
action, giving an end-point dependence to the action to higher orders.

Because $\delta\phi(\tau_0)=0$, we have
\ba
\delta\phi(\tau)&&=\int_{\tau}^{\tau_0}\delta\phi' d\tilde\tau\\
\delta\phi'&&= -{1\over a^{(0)3}}\int_0^\tau a^{(0)3} d\tilde\tau
\ea
Because of the $1/a^3$ term in the integrand for $\delta \phi$, its 
value   will depend, among other things,
   on the way in which the path in $\tau $ space wraps
around the poles where $a=0$. We get
\beq
\delta\phi(\tau)={2\over V_0}\left[ {1\over 1+\cos(i\nu\tau)}
 +\ln\left({1+\cos(i\nu\tau)\over 1+\cos(i\nu\tau_0)}\right) 
 - {1\over 1+\cos(i\nu\tau_0)}\right]
\eeq
where the branches of the ln function around  the singularities where
$1+\cos(i\nu\tau)=0$ are chosen so as to make the expression
$\delta\phi(\tau(t))$
continuous along the path $\tau(t)$.

\subsection{Second Order Action}
Given the first order solutions, we can calculate the action to second order in
$\epsilon$. 
\beq
\delta^2S = \phi_f^2{d^2\over {dV_0}^2} S_0(a_f,\phi_f,V_0) - \int_0^{\tau_0}
a^{(0)3}(\delta\phi')^2 d\tau
\eeq
 We have
 \ba
 -\int_0^{\tau_0} a^{(0)3}(\delta\phi')^2 d\tau&&= \beta\int_0^{\tau_0} 
 + {1 \over \nu^5\sin^3(i\nu\tau)} \left({1\over 3}
\cos^3(i\nu\tau) -\cos(i\nu\tau)+{2\over 3}\right)^2 d\tau\\
&&=+{i\beta\over 9\nu^6}\int_1^{cos(i\nu\tau_0)} 
\left( (1-z)(z+2)\over 1+z\right)^2 dz
\ea
with $z=cos(i\nu\tau)$
which has poles of order 2 in the integrand at
at $z=-1$ or $\tau=i(2m+1)\pi/\nu$. 
Each time the integrand circles one of these poles, we accumulate a residue, and
it is clear that the residue is the same at each of the poles in $\tau$ space,
since it is the same pole $z=-1$ in z space. Thus we have
\beq
-\int_0^{\tau_0} a^{(0)3}(\delta\phi')^2 d\tau=
+\beta {i\over 9\nu^6}\int_{\Gamma_0} \left( (1-z)(z+2)\over 1+z\right)^2 dz  +
2\pi \beta k {96\over
V_0^3}
\eeq
where $\Gamma_0$ is some fiducial path from the point $\tau=0$ to $\tau=\tau_0$,
and $k$ is the total net number of times that the actual path wraps around the
 various poles. (For an encircling of each of the poles in the 
 counter clockwise
 direction, one gets the same contribution no matter which of the poles is
 chosen.)  Note that the contribution from each of the poles to the action is
 real. They will therefore contribute only to the phase, and not the amplitude,
 of the wave-function.
 
 Assuming that $\nu a_f>1$, we can take the fiducial path as a straight
 line connecting $\tau=0$ to $\tau=\tau_f$. We finally get
 \ba
 \delta^2S= \beta&&\left[\phi_f^2 {\partial^2
 S_0(a_f,\phi_f,V_0)\over \partial V_0^2}  -\pi k {192\over V_0^3}\right.\\ 
 &&\left. +{8i\over V_0^3 }\left[  -{12\over Z+1} +12 \ln\left({Z+1\over 2}\right)
  +{17} -12Z + Z^3\right]
 \right]
 \ea
 where $Z= cos(i\nu\tau_0) = \gamma i\vert \sqrt{{V_0 a_f^2\over 6}-1}\vert$.
 
 We note that once again, the action does not depend on the endpoint of the
 integral--ie on which of the end values of $\tau_0$ the integrand finishes on.
  
  We find that $\delta^2S$ also obeys the second order Hamilton Jacobi
  function for the Einstein action. However it is important to note that to
  second order, $S$ is a complex function of $a_f,\phi_f$, with the imaginary
  part an actual function of these variables, instead of being a constant (as it
  is to zeroth order). Since the Hamilton Jacobi equation is a non-linear
  equation to second order in $\epsilon$, the real part of $S$ will not be a
  Hamilton Jacobi function for the Einstein action.  This will be important in
  the next section where we look at the classical paths which the wave-functions
  for the universe imply.
  
  \section{Classical Evolution}
  
  The impression is often conveyed that the semi-classical path integral implies
  that the history that the universe follows is just the  classical solution
  which is used to dominate the path integral. It is clear that this cannot be
  the correct answer. We have seen that the wave-function of the universe has
  contributions not just from real solutions to Einstein's equations, but also
  from complex solutions. These complex solutions clearly cannot represent a
  history of the universe, since physical things like the radius of the
  universe, the value of the scalar field, and their velocities in (proper) time
  are real quantities, not complex quantities. The semi-classical approximation
  to the path integral is a mathematical trick for evaluating the wave function,
  and does not in itself represent anything ``real", and in particular does not
  represent a real history. However, once one has a wave function, one can ask
  the quantum mechanical question as to what types of history of the universe 
  that wave function  will represent. This is a very deep problem which does not
  have any satisfactory solution as yet. However, this paper is not the place
  for a detailed examination of the various possibilities. Instead we will take
  a naive approach.
  
  In ordinary quantum mechanics, the momentum operator conjugate to the position is
  just the imaginary derivative of the wave function. Although this operator
  really only makes sense if integrated over the configuration space, we will
  assume that some sense can be made of the momentum density, $\psi(q)^*
  i{\partial\over\partial q} \psi(x)$. However this quantity is not real, while a
  physical momentum should be real. Furthermore, the amplitude squared of the
  wave function, $|\psi(q)|^2 $ is the probability of finding the particle at
  the position $q$. The derivative of the amplitude of the wave function is thus
  not really something associated with the momentum of the particle. We can
  therefore take as the definition of the momentum at the point x, the quantity
  $\tilde\psi(q) i{{\partial\over\partial q}\tilde\psi(q)}$, where
  $\tilde\psi(q)=\psi(q)/|\psi(q)|$. Thus the momentum is just the derivative of
  the phase of the wave function. If the wave function is of the form $e^{iS}$,
  then the momentum will be just the derivative of the {\bf real} part of the
  action $S$. This argument is clearly highly suspect, and is at best only a
  crude approximation. However, it is the approximation which we shall use.
  Thus, the Real part of the action $S$ will be used as a Hamilton Jacobi
  function for the evolution of the system as an approximation to the classical
  evolution implied by the quantum wave function. Again, we emphasize most
  strongly that this is at best a very crude attempt at deriving a classical
  evolution from the quantum system. In particular, in cases of significant
  quantum interference, this approximation will break down entirely. (for
  example, for a purely real wave-function, this gives the result
  that the momentum is everywhere zero).
  
  The lowest order action is complex, but the imaginary terms do not depend on
  either $a_f$ or $\phi_f$.  In the first order action, the imaginary part does
  depend on $\phi_f$, but since the zeroth order action is entirely independent
  of $\phi_f$, and the Hamilton Jacobi equation has $\partial S\over \phi_f$ to
  the second order, this derivative of the imaginary part of the action will not
  contribute to the first order HJ equation. Thus, the real part of the first
  order action
  will again obey the HJ equation to first order. However, to second order, this
  is no longer true. The imaginary part of the first order action depends on
  $\phi_f$ and thus comes in in quadratic order in the second order HJ equation.
  Ie, the real part of the second order action will not be a solution of the HJ
  equation, even though the full complex second order action is.
   Thus, the histories obtained by using only the real part of the
  action will not be solutions of the Einstein equations. Einstein's equations
  will have ``quantum corrections" arising from the HH specification for the
  wave-function of the universe.
  
  In particular, if we use the real part of our second order solution, it obeys
  the HJ equation to second order in epsilon of
  \ba
  &&-{1\over 12a}\left[\left({\partial \delta S_{R}\over \partial a}\right)^2 + 2{\partial S_0\over
  \partial a}  {\partial \delta^2 S_R\over \partial a}\right] + {1\over
  a^3}\left({\partial \delta S_R\over \partial \phi}\right)^2\\
    && = -{1\over
  12a}\left({\partial\delta S_I\over\partial a}\right)^2 
  +{1\over a^3} \left({\partial \delta S_I\over \partial \phi}\right)^2 
  \ea
  But in order for $S_R$ to be a solution of the HJ equation for
   Einstein's equations to second order in $\epsilon$, the
  right hand side would have to be zero. Instead it is $576\over V_0^4a^3$. We
  could subtract a term from the Einstein Hamiltonian constraint to find a
  new Hamiltonian which the real part would satisfy.  We note that this term
  represents a potential which diverges as $a$ goes to zero, a potential which
  will help prevent the universe from collapsing.
  \beq
  H_{mod}= -{\pi_a^2\over 24a} - 6a +{\pi_\phi^2\over 2a^3 } +a^3 V(\phi) -288
  {\epsilon^2\over V_0^4a^3}
  \eeq
  Note that the ``quantum" correction falls off rapidly with $a$, which implies
  that once the universe is large, the impact of the imaginary parts of the
  action on the evolution of the universe become negligible.
  
  Let us examine the classical solutions obtained by using $S_R$ as a
   HJ function for the solution. We are interested in the motion for $\nu
   |a_f|>1$, Furthermore, we expect $V_0$ to enter into the complete action only
   in the combination $V=V(\phi_f)=V_0+\epsilon \phi_f$. Thus, we define 
   \beq
   y=\vert \sqrt{{V a_f^2\over 6} -1}\vert
   \eeq
   and have
  \ba
  S_R\approx +&&\beta\gamma{24\over V} y^{3}  \nonumber\\
  &&+\epsilon^2\beta{4\over V^3 }  \left( -12{\gamma y\over 1+ y^2} +6i
  \ln{1+i\gamma y \over 1-i\gamma y}  + 12\gamma y +
  \gamma y^3 -24 \pi k\right)   
  \ea
  (Note that the term, $6i  \ln{1+i\gamma y \over 1-i\gamma y}$ could also be
  written as $-12\gamma arctan(  y)$ which makes it clear that is is real).
  The equations relating the derivatives with respect to time 
  of the dynamic variables  
  and the HJ function are
  \ba
  \pi_a= -12a \dot a={\partial S_R\over \partial a} \\
  \pi_\phi= a^3\dot\phi={\partial S_R\over \partial \phi}
  \ea
  where we have chosen the time to be the proper time in the space-time (ie,
  N=1).
  
  For large $a$,  we have 
  \beq
  S_R\approx \beta\gamma |a|^3  4 \sqrt{V(\phi)\over 6}\left[ 1 +{1\over 6 V^2}\epsilon^2 \right]
  \eeq
  and the evolution of $\phi\approx 2\epsilon\sqrt{6\over V_0} t$, while $a$
  grows exponentially. These are just the ``slow roll" results for the dynamics
  in an inflationary system. $\epsilon$ makes only a small change in the results.
  
  What is more interesting is the behaviour of $a$ and $\phi$ near the minimum,
  when $V(\phi)a^2/6 $ is of order unity. We first note that we can rewrite the
  expression for $S_R$ as
  \beq
  S_R\approx \beta\gamma{24\over V}\left( (1 +\epsilon^2{17\over
  6V^2})(y^2-\epsilon^2{4\over 3V^2})^{3\over 2}\right)
   +\epsilon^2 O(y^5) +O(\epsilon^3)
   \eeq
   Ie, to second order in $\epsilon$ and to fourth order in $y$ this expression
   is identical to the previous expression for $S_R$, as can be readily verified
   by expanding this expression to second order in $\epsilon$.
   Since
   \beq
    y^2-\epsilon^2{4\over 3V^2}= {V_0\over 6}a^2 -1 -\epsilon^2{4\over 3V^2}
    +\epsilon{1\over 6}\phi a^2
     \approx {V_0\over 6}(a-a_T)(2a_T) +\epsilon{1\over 6} \phi a_T^2 +O((a-a_T)^2)
   \eeq
   where 
   \beq
   a_T= \sqrt{6\over V_0} (1+\epsilon^2 {2\over 3V_0^2})
   \eeq
   we can write the action as 
   \beq
   S_R\approx \beta\gamma {24\over V} \left( (1+ {17\over 6V^2}\epsilon^2  )
    \left(
    {V_0a_T\over 3} (|a|-a_T) +\epsilon {a_T^2\over 6}\phi \right)^{3\over 2}\right)
   \eeq
   Defining $\xi= |a|-a_T$, we have
   \ba
   \pi_a=-12a \dot a \approx -12 a_T \dot\xi&&\approx \beta\gamma {24\over
   V_0}(1+\epsilon^2{17\over 6V_0^2}) {3\over 2}{V_0
   a_T\over 3} ({V_0a_T\over 3}\xi+\epsilon{a_T^2\over 6}\phi)^\half\\
   &&\approx \beta\gamma 12a_T (1+\epsilon^2{17\over 6V_0^2})
   ({V_0a_T\over 3}\xi+\epsilon{a_T^2\over 6}\phi)^\half\\
   \pi_\phi=a_T^3 \dot\phi &&\approx
    \epsilon\beta\gamma {24\over
   V_0} 
   {3\over 2}{a_T^2\over 6}  ({V_0a_T\over 3}\xi+\epsilon{a_T^2\over 6}\phi)^\half\\
   &&\approx \beta\gamma\epsilon {6a_T^2\over V_0}
   ({V_0a_T\over 3}\xi+\epsilon{a_T^2\over 6}\phi)^\half
\ea
where we have only kept terms to lowest order in $\xi$ and $\phi$. 
Solving these equations we find that 
\beq
({V_0a_T\over 3}\xi+\epsilon{a_T^2\over 6}\phi)^\half
= -\beta\gamma{V_0 a_T\over 6}(1+\epsilon^2{35 \over 6 V_0^2})
    t
\eeq
Since $({V_0a_T\over 3}\xi+\epsilon{1\over 6}\phi)^\half$ was taken as being
positive, we have that $sign(t) = -\beta\gamma$. Ie, the choice of $\beta$ and
$\gamma$ determines whether the solution starts or ends and that minimal value
of $a=a_T$. We also note that $\epsilon$ only causes small (or order $\epsilon^2$)
changes in the  solution.

We finally have
\ba
|a|-a_T\approx (\half a_T +O(\epsilon^2)) t^2\\
\phi\approx -\epsilon\half t^2
\ea

Note that the main effect of the HH boundary conditions has been to demand that
$\phi$ is a maximum in time at the time when $a$ is a minimum. One might have
expected $\dot\phi$ to be non-zero at this minimum, but it is not.  This means that for these
states, the value of $\phi$ cannot be used a clock near the turn around point, since the clock
changes sign at point, although in principle $\pi_\phi$ could be.

Of course, this turning point is also where one would expect to find maximal
quantum interference between the various possible semi-classical paths, and so
this non-analytic behaviour of the classical solution is suspect. However we
note that the semi-classical wave-function, if we take an equal   linear combination
of terms with $\gamma=\pm 1$, can be constantly interpreted not as a wave
function which emerges out of the $a=0$ ``singularity" (that would be consistent
with taking only the solution with $\gamma=-\beta$), but rather with the
boundary condition imposing the condition that the classical universe suffers a
bounce. Ie, the HH conditions could be interpreted as enforcing a bouncing
universe, rather than as describing the creation of the universe out of nothing.

We note that we have neglected the contribution of the fluctuations to the path
integral in all of the above analysis. One would expect that these too would be
analytic in the path through $\tau$ space which one chooses. However, these
fluctuation terms could depend on which of the ``end points" the path in $\tau$
space ended at. In fact, because the first order solution $\delta a(\tau)$
 is not periodic in imaginary $\tau$, one would expect the higher order actions
 to also depend on which of these end-points was chosen in defining the
 semi-classical action. Both the fluctuations and the end-point dependence could 
  could substantially change the behavior of the wave function,
especially near the ``turning point".  

\section{Conclusion}
We have solved the equations for the evolution of the universe in an
inflationary approximation, where the slope of the potential driving the scalar
field can be taken to be small. There exist an infinite number of complex
solutions which can be said to dominate the path integral. Although some of
these paths are analytically related to each other, others are not. The analytically related 
paths, which all have the same action, could, as others have done, be regarded as equivalent in some sense. Since the manifold
is real they are not related by a diffeomorphism, and the question as to whether or not they
should be summed over or only one representative of the class should be used in defining the
wave-function is to us still an open question. However,
 there exist singular points for the solutions in the complex $\tau$ plane, which
correspond to $a$ going to zero. The path in the complex plane can wrap around
these points, and the action depends on how the paths wrap around these singularities. 
To the order of approximation we examine, these singularities contribute to the real part of the path
integral to second order in the slope of the potential. There exist an infinite number of these
inequivalent analytic paths.  Furthermore,
there are also an infinite variety of analytically inequivalent paths which depend on the
choice of the end point of the integration in the complex $\tau$ plane. 
 The action, to second order, depends only on which of two classes the ends points fall into (as
designated by the the parameter $\gamma$ above), although there are indications that to higher
order the action will also depend on exactly which of these end points is chosen.
  
  If one assumes that the classical path that a universe would follow is best
  approximated by using the real part of the semi-classical complex action as
  the Hamilton Jacobi function, then one interpretation of the Hartle-Hawking
  wave function is not that the universe is created out of nothing, but rather
  that the HH condition forces the condition that the universe bounce rather
  than enter or leave the potential singularity. Of course it is precisely at
  the bounce point that the role of the quantum corrections (here taken to be
  the imaginary part of the semi-classical action) have their largest value, and
  the interpretation of the universe following a classical path is most
  problematic. 
  
  We also note that our analysis demonstrates that , in the absence of
  non-homogeneous modes, the HH condition does not select the sign of the real
  part of the action (the parameter $\beta$ above) and thus this analysis does
  not have anything to say to the controversy over this sign. This model however
  seems simple enough that it could also be used to address the issue of
  non-homogeneous fluctuations, at least to some low order of approximation,
  which could cast some light on the controversy. However, we believe that this
  would require a deeper understanding than we at least have of the role that
  such complex solutions really play in the evaluation of a path integral.
  
 What has not been done here is to calculate the contribution of the "determinant", (the
integration over non-classical fluctuations), to the wave function. It is possible that this
would also lift the degeneracy over the various possible paths. This paper also does not address
the broader question of the how one should use such complex solutions in evaluating the path
integral. However, even in the context of ordinary quantum mechanics with a single degree of
freedom, this question of how
complex classical paths are to be used in general in evaluating the propagator  is still one
requiring more understanding.

\section*{Acknowledgements}
WGU would like to thank NSERC and the Canadian Institute for Advanced Research
for support while this work was carried out.

\end{document}